\documentstyle[multicol,aps,prl,epsf]{revtex}
\begin{document}
\def\bea{\begin{eqnarray}}
\def\eea{\end{eqnarray}}
\title{Persistence in Nonequilibrium Systems}
\author{Satya N. Majumdar\\
{Theoretical Physics Group, Tata
Institute of Fundamental Research, 
\\Homi Bhabha Road, Bombay 400005, India. \\}
}
\date{\today}
\maketitle

\begin{abstract}
This is a brief review of recent theoretical efforts to understand  
persistence in nonequilibrium systems. Some of the recent experimental
results are also briefly mentioned. I also discuss recent generalizations
of persistence in various directions and conclude with a summary of open
questions. 

\end{abstract}

\begin{multicols}{2}
The problem of {\it persistence} in spatially extended nonequilibrium
systems has recently generated a lot of interest both
theoretically\cite{derrida1,derrida2,bray,majsire,diffusion,majbray,watson}
and experimentally\cite{marcos,yurke,tam}. Persistence is simply the
probability that the fluctuating nonequilibrium field does not change sign
upto time $t$. 
These systems include various models undergoing
phase separation 
process\cite{derrida1,derrida2,bray,majsire,crit,clement,lee,krapiv,majcor},
simple diffusion equation with random
initial conditions\cite{diffusion,majbray}, several reaction diffusion
systems in both pure\cite{cardy} and disordered\cite{fisher} environments, 
fluctuating interfaces\cite{krug,kallabis,tnd},
Lotka-Volterra models of population dynamics\cite{frache} and
granular medium\cite{swift}. 

The precise definition of persistence is as
follows. Let $\phi(x,t)$ be a
nonequilibrium field fluctuating in space and time according to some
dynamics. For example, it could
represent the coarsening spin field in the Ising model after being
quenched to low temperature from
an initial high temperature. It could also be
simply a diffusing field starting from random initial
configuration or the height of a fluctuating
interface. Persistence is simply the probability $P_0(t)$ that at a fixed
point in space, the quantity $sgn[\phi(x,t)-\langle \phi(x,t)\rangle]$
does not change upto time $t$. In all the examples mentioned above this
probability decays as a power law $P_0(t)\sim t^{-\theta}$ at late times,
where the persistence exponent $\theta$ is usually nontrivial.
 
In this article, we review some recent theoretical efforts in calculating
this nontrivial exponent in various models and also
mention some recent experiments that measured this exponent.
The plan of
the paper is as follows. We first discuss the persistence in
very simple single variable systems. This makes the ground for later
study of persistence in more complex many body systems. Next we consider
many body systems such as Ising model and discuss where the complexity is
coming from. We follow it up with the calculation of this exponent for a
simpler many body system namely diffusion equation and see that even in
this simple case, the exponent $\theta$ is nontrivial. Next we show that
all these examples can be viewed within the general framework of the 
``zero crossing" problem of a Gaussian stationary process (GSP). We review
the new results obtained for this general Gaussian problem in various
special cases. Finally we mention the emerging new directions towards
different generalizations of persistence.

We start with a very simple system namely the one
dimensional Brownian walker. Let $\phi(t)$ represent the position of a
$1$-d Brownian walker at time $t$. This is a single body system in the
sense that the field $\phi$ has no $x$ dependence but only $t$ dependence.
The position of the walker evolves as,
\bea
\frac{d\phi}{dt}=\eta(t)  
\label{brown}
\eea
where $\eta(t)$ is a white noise with zero mean and delta correlated,
$\langle \phi(t)\phi(t')\rangle=\delta(t-t')$. Then persistence $P_0(t)$
is
simply the
probability that $\phi(t)$ does not change sign upto time $t$, i.e,
the
walker does not cross the origin upto time $t$. This problem can be very
easily solved exactly by writing down the corresponding Fokker-Planck
equation
with an absorbing boundary condition at the origin\cite{feller}. The
persistence
decays as $P_0(t)\sim t^{-1/2}$ and hence $\theta=1/2$. The important
point
to note here is that the exact calculation is possible here due to the
Markovian nature of the process in Eq. (1). Note that $\phi$ evolves
according to a first order equation in time, i.e., to know $\phi(t)$, we
just need the value of $\phi(t-\Delta t)$ but not on the previous history.
This is precisely the definition of a Markov process.
 
In order to make contact with the general framework to be developed in
this article, we now solve the same process by a
different method. We note from Eq. (1) that $\eta(t)$ is a
Gaussian noise and Eq. (1) is linear in $\phi$. Hence $\phi$ is also a
Gaussian process with zero mean and a two time correlator, $\langle
\phi(t)\phi(t')\rangle=min(t,t')$ obtained by integrating Eq. (1).
We recall that a Gaussian process can be completely characterized by just
the two-time correlator. Any higher order correlator can be simply
calculated by using Wick's theorem. Since $min(t,t')$ depends on both
time $t$ and $t'$ and not just on their difference $|t-t'|$, clearly
$\phi$ is a Gaussian non-stationary process. From the technical point
of view, stationary processes are
often preferable to non-stationary processes. Fortunately there turns out
to be a simple transformation by which one
can convert this non-stationary process into a stationary one. It turns
out that this transformation is more general and will work even for more 
complicated examples to follow. Therefore we
illustrate it in detail for the Brownian walker problem
in the following paragraph.

The transformation works as follows. Consider first the normalized
process, $X(t)=\phi(t)/{\sqrt {\langle \phi^2(t)\rangle}}$. Then $X(t)$
is also a Gaussian process with zero mean and its two-time correlator is
given by, $\langle X(t)X(t')\rangle =min(t,t')/{\sqrt {(tt')}}$. Now we
define a new ``time" variable, $T=\log (t)$. Then, in this new time
variable $T$, the two-time correlator becomes, $\langle X(T)X(T')\rangle
=\exp (-|T-T'|/2)$ and hence is stationary in $T$.
Thus, the persistence problem reduces to calculating the probability
$P_0(T)$ of no zero crossing of $X(T)$, a
GSP characterized by its two-time
correlator, $\langle X(T)X(T')\rangle=\exp(-|T-T'|/2)$.

One could, of course, ask the same question for an arbitrary GSP with a
given correlator $\langle X(T)X(T')\rangle
=f(|T-T'|)$ [in case of Brownian motion, $f(T)=\exp(-T/2)$]. This
general zero crossing problem of a GSP
has been studied by mathematicians for a long time\cite{slepian}.
Few results are known exactly. For example, it is known that if $f(T)<1/T$
for large $T$, then $P_0(T)\sim \exp(-\mu T)$ for large $T$. Exact result
is known only for Markov GSP which are
characterized by purely exponential correlator, $f(T)=\exp(-\lambda T)$.
In that case, $P_0(T)={2\over {\pi}}{\sin}^{-1}[\exp(-\lambda
T)]$\cite{slepian}.
Our example of Brownian motion corresponds to the case when $\lambda=1/2$
and therefore the persistence $P_0(T)\sim \exp(-T/2)$ for large $T$.
Reverting back to the original time using $T=\log (t)$, we recover the
result, $P_0(t)\sim t^{-1/2}$. Thus the inverse of the decay rate in 
$T$ becomes the power law exponent in $t$ by virtue of this
``log-time" transformation. Note that when the correlator $f(T)$ is
different from pure exponential, the process is non-Markovian and in that
case no general answer is known.

Having described the simplest one body Markov process, we now consider
another one body process which however is non-Markovian. Let $\phi(t)$
(still independent of $x$) now represents the position of a particle
undergoing random acceleration,
\bea
\frac{d^2\phi}{dt^2}=\eta(t)  
\label{accl}
\eea
where $\eta(t)$ is a white noise as before. What is the probability
$P_0(t)$
that the particle does not cross zero upto time $t$? This problem was
first proposed in the review article by Wang and Uhlenbeck\cite{wang} way
back in 1945 and it got solved only very recently in 1992, first by
Sinai\cite{sinai}, followed by Burkhardt\cite{burkhardt} by a different
method. The answer is, $P_0(t)\sim t^{-1/4}$ for large $t$ and the
persistence exponent is $\theta=1/4$. Thus even for this apparently simple
looking problem, the calculation of $\theta$ is nontrivial. This
nontriviality can be traced back to the fact that this process is
non-Markovian. Note that Eq. (2) is a second order equation and to know
$\phi(t+\Delta t)$, we need to know its values at two previous points
$\phi(t)$ and $\phi (t-\Delta t)$. Thus it depends on two previous steps
as opposed to just the previous step as in Eq. (1). Hence it is a
non-Markovian process. 

We notice that the Eq. (2) is still linear and
hence $\phi(t)$ is still a Gaussian process with a non-stationary
correlator. However, using the same $T=\log (t)$ transformation as defined
in the previous paragraph, we can convert this to the zero crossing
problem in time $T$ of a GSP with correlator,
$f(T)={3\over {2}}\exp(-T/2) -{1\over {2}}\exp(-3T/2)$. Note that this
is different from pure exponential and hence is non-Markovian. We also
notice another important point: It is not correct to just consider the
asymptotic form of $f(T)\sim {3\over {2}}\exp(-T/2)$ and conclude that
the exponent is therefore $1/2$. The fact that the exponent is exactly
$1/4$, reflects that the ``no zero crossing" probability $P_0(T)$ depends
very crucially on the full functional form of $f(T)$ and not just on its
asymptotic form. This example thus illustrates the history dependence
of the exponent $\theta$ which makes its calculation nontrivial. 

Having discussed the single particle system, let us now turn to many body
systems where the field $\phi(x,t)$ now has $x$ dependence also. The first
example that were studied is when $\phi(x,t)$ represents the spin field
of one dimensional Ising model undergoing zero temperature coarsening
dynamics, starting from a random high temperature configuration. Let us
consider for simplicity a discrete lattice where $\phi(i,t)=\pm 1$
representing Ising spins. One starts from a random initial
configuration of these spins. The zero temperature dynamics
proceeds as follows: at every step, a spin is chosen at random and 
its value is updated to that of one of its neighbours chosen at random and
then time is incremented by $\Delta t$ and one keeps repeating this
process. Then persistence is simply the probability that a given spin (say
at site $i$) does not flip upto time $t$. Even in one dimension, the
calculation of $P_0(t)$ is quite nontrivial. Derrida et.
al.\cite{derrida2} solved this problem exactly and found $P_0(t)\sim
t^{-\theta}$ for large $t$ with $\theta=3/8$. They also 
generalized this to $q$-state Potts model in $1$-d and found an exact
formula, $\theta(q)=-{1\over {8}}+{2\over
{\pi^2}}[\cos^{-1}\{(2-q)/q{\sqrt {2}}\}]^2$ for all $q$. 

This calculation however can not be easily extended to $d=2$ which is more
relevant from experimental point of view. Early numerical results
indicated that the exponent $\theta\sim 0.22$\cite{bray} for $d=2$ Ising
model evolving with zero temperature spin flip dynamics. It was therefore
important to have a theory in $d=2$ which, if not exact, at least could
give approximate results. We will discuss later about our efforts towards
such an approximate theory of Ising model in higher dimensions. But before
that let us try to understand the main difficulties that one encounters in
general in many body systems.

In a many body system, if one sits at a particular point $x$ in space and
monitors the local field $\phi (x,t)$ there as a function of $t$, how
would this ``effective" stochastic process (as a function of time only)
look like?
If one knows enough properties of this single site process as a function
of time, then the next step is to ask what is the probability that this
stochastic process viewed from $x$ as a function of $t$, does not change
sign upto time $t$. So the general strategy involves two steps: first, one
has to solve the underlying many body dynamics to find out what the    
``effective" single site process looks like and second, given this single
site process, what is its no zero crossing probability. 

Before discussing
the higher dimensional Ising model where both of these steps are quite
hard, let us discuss a simple example (which however is quite abundant in
nature) namely the diffusion equation. This is a many body system but
at least the first step of the two-step strategy can be carried out
exactly
and quite simply. The second step can not be carried out exactly even for
this simple example, but one can obtain very good approximate results.

Let $\phi(x,t)$ (which depends on both $x$ and $t$) denote field that is
evolving via the simple diffusion equation,
\bea
\frac{\partial \phi}{\partial t}=\nabla^2 \phi.  
\label{diff}
\eea
This equation is deterministic and the only randomness is in the initial
condition $\phi(x,0)$ which can be chosen as a Gaussian random variable
with zero mean. For example, $\phi(x,t)$ could simply represent the
density fluctuation, $\phi(x,t)=\rho(x,t)-\langle \rho\rangle$ of a
diffusing gas. The persistence, as usual, is simply the probability that
$\phi (x,t)$ at some $x$ does not change sign upto time $t$. This
classical diffusion equation is so simple that it came as a surprise to
find that even in this case, the persistence $P_0(t)\sim t^{-\theta}$
numerically with nontrivial $\theta\approx 0.1207$, $0.1875$, $0.2380$ in
$d=1$, $2$ and $3$ respectively.

In light of our previous discussion, it is however easy to see why one
would expect nontrivial answer even in this simple case. Since the
diffusion equation (3) is linear, the field $\phi(x,t)$ at a fixed point
$x$ as a function of $t$ is clearly a Gaussian process with zero mean
and is simply given by the solution of Eq. (3), 
$\phi(x,t)=\int d^dx' G(\vec x -\vec x',t)\phi(x',0)$, where
$G(\vec x,t)=(4\pi t)^{-d/2}\exp [-x^2/{4t}]$ is the Green's function in
$d$. Note that by solving the Eq. (3), we have already reduced the many
body diffusion problem to an ``effective" single site Gaussian process in
time $t$ at fixed $x$. This therefore completes the first step of the 
two-step strategy mentioned earlier exactly. Now we turn to the second
step, namely the ``no zero crossing" probability of this single site
Gaussian process. The two time correlator of this can be easily computed
from above and turns out to be non-stationary as in the examples of
equations (1) and
(2). However by using the
$T=\log(t)$ transformation as before, the normalized field reduces to a
GSP in time $T$ with correlator, $ \langle
X(T_1)X(T_2)\rangle=[{\rm {sech}}(T/2)]^{d/2}$, where $T=|T_1-T_2|$.
Thus once again, we are back to the general problem of the zero crossing
of a GSP, this time with a correlator
$f(T)=[{\rm {sech}}
(T/2)]^{d/2}$ which is very different from pure exponential form and hence
is non Markovian. The persistence, $P_0(T)$ will still decay as
$P_0(T)\sim
\exp(-\theta T)\sim t^{-\theta}$ for large $T$ (since $f(T)$ decays faster
than $1/T$ for large $T$) but clearly with a nontrivial exponent.

Since persistence in all the examples that we have
discussed so far (except the Ising model) reduces to the zero crossing
probability of a GSP with correlator
$f(T)$ [where $f(T)$ of course varies from problem to problem], let us now
discuss some general properties of
such a process. It turns out that a lot of information can already be
inferred by examining the short-time properties of the correlator $f(T)$.
In case of Brownian motion, we found $f(T)=\exp(-T/2)\sim 1-T/2 +O(T^2)$
for small $T$. For the random acceleration problem,
$f(T)={3\over {2}}\exp(-T/2) -{1\over {2}}\exp(-3T/2)\sim 1-3T^2/8
+O(T^3)$ for small $T$ and for the diffusion problem, $f(T)=[{\rm {sech}}
(T/2)]^{d/2}\sim 1-{d\over {16}}T^2 +O(T^3)$ as $T\to 0$. In general
$f(T)=1-aT^{\alpha}+\ldots $ for small $T$, where $ 0<\alpha \leq
2$\cite{slepian}. It turns out that processes for which $\alpha=2$ are
``smooth" in the sense that the density of zero crossings $\rho$ is
finite, i.e., the number of zero crossings of the process in a given
time $T$ scales linearly with $T$. Indeed there exists an exact formula
due
to Rice\cite{rice}, $\rho={\sqrt {-f''(0)}}/{\pi}$ when $\alpha=2$.
However, for $\alpha<2$, $f''(0)$ does not exist and this formula
breaks down. It turns out that the density is
infinite for $\alpha<2$ and once the process crosses zero, it immediately
crosses many
times and then makes a long excursion before crossing the zero again. In
other words, the zero's are not uniformly distributed over a given
interval and in general the set of zeros has a fractal
structure\cite{reviews}.

Let us first consider ``smooth" processes with $\alpha=2$ such as
random acceleration or the diffusion problem. It turns out that for
such processes, one can 
make very good progress in calculating the persistence exponent $\theta$.

The first approach consists of using an "independent interval
approximation" (IIA)\cite{diffusion}. Consider the ``effective" single
site process $\phi(T)$ as a function of the ``log-time" $T=\log (t)$. As a
first step, one introduces the
``clipped" variable $\sigma=sgn(\phi)$, which changes sign at the zeros of 
$\phi(T)$. Given that $\phi(T)$ is a Gaussian process, it is easy to
compute the correlator, $A(T)=\langle
\sigma(0)\sigma(T)\rangle={2\over {\pi}}{\sin}^{-1}[f(T)]$, where $f(T)$
is the correlator of $\phi(T)$. Since the ``clipped" process $\sigma(T)$
can take values $\pm 1$ only, one can express $A(T)$ as,
\bea
A(T)=\sum_{n=0}^{\infty} (-1)^n P_n(T),     
\label{clipped}
\eea
where $P_n(T)$ is the probability that the interval $T$ contains $n$ zeros
of $\phi(T)$. So far, there is no approximation. The strategy next is to
use the following approximation,
\bea
P_n(T)&=&{\langle
T\rangle}^{-1}\int_{0}^TdT_1\int_{T_1}^{T_2}dT_2\ldots\int_{T_{n-1}}^{T}dT_n
\nonumber \\
&\times& Q(T_1)P(T_2-T_1)\ldots P(T_n-T_{n-1})Q(T-T_n),
\label{iia}
\eea
where $P(T)$ is the distribution of intervals between two successive
zeros
and $Q(T)$ is the probability that an interval of size $T$ to the right or
left of a zero contains no further zeros. Clearly, $P(T)=-Q'(T)$.
$\langle
T\rangle=1/\rho$ is the mean interval size. We have made the IIA by
writing the joint distribution of $n$ successive zero-crossing intervals
as the product of the distribution of single intervals. The rest is
straightforward\cite{diffusion}. By taking the Laplace transform 
of the above equations, one finally obtains, ${\tilde
P}(s)=[2-F(s)]/F(s)$
where,
\bea
F(s)= 1+ {1\over {2\rho}}s[1-s {\tilde A}(s)],
\label{zero}
\eea
where the Laplace transform ${\tilde A} (s)$ of $A(T)$ can be easily
computed knowing $f(T)$. The expectation that the persistence, $P_0(T)$
and hence the interval distribution, $P(T)\sim \exp(-\theta T)$ for large
$T$, suggests a simple pole in the ${\tilde P}(s)$ at $s=-\theta$. The
exponent
$\theta$ is therefore given by the first zero on the negative $s$ axis of
the function,
\bea
F(s)=1+{1\over {2\rho}}s\big\{1-{2s\over {\pi}}\int_0^{\infty}dT \exp
(-sT){\sin}^{-1}[f(T)]\big\}.
\label{final}
\eea

For the diffusion equation, $f(T)=[{\rm {sech}}(T/2)]^{d/2}$ and $\rho=\sqrt
{d/{8\pi^2}}$. We then get the IIA estimates of $\theta=0.1203$, $0.1862$
and $0.2358$ in $d=1$, $2$ and $3$ respectively which should be compared
to the simulation values, $0.1207 \pm 0.0005$, $0.1875\pm 0.0010$ and
$0.2380\pm 0.0015$. For the random acceleration problem, 
$f(T)={3\over {2}}\exp(-T/2) -{1\over {2}}\exp(-3T/2)$
and $\rho=\sqrt{3}/{2\pi}$ and we get, $\theta_{iia}=0.2647$ which can be
compared to its exact value, $\theta=1/4$. 

Note that the IIA approach, though it produces excellent results when
compared to numerical simulations, cannot however be systematically
improved. For this purpose, we turn to the ``series expansion" 
approach\cite{majbray} which can be improved systematically order by
order. The idea is to consider the generating function,
\bea
P(p,t)=\sum_{n=0}^{\infty} p^n P_n(t)
\label{genfunc}
\eea
where $P_n(t)$ is the probability of $n$ zero crossings in time $t$ of the
``effective" single site process. For $p=0$, $P(0,t)$ is the usual
persistence, decaying as $t^{-\theta(0)}$ as usual. Note that we have used
$\theta(0)$ instead of the usual notation $\theta$, because it turns
out\cite{majbray}
that for general $p$, $P(p,t)\sim t^{-\theta (p)}$ for large $t$, where
$\theta(p)$ depends continuously on $p$ for ``smooth" Gaussian processes.   
This has been checked numerically as well as within IIA
approach\cite{majbray}. Note that for $p=1$, $P(1,t)=1$ implying
$\theta(1)=0$. For smooth Gaussian processes, one can then derive an exact
series expansion of $\theta(p)$ near $p=1$. 
Writing $p^n=\exp (n\log p)$ and expanding the
exponential, we then obtain an expansion in terms of of moments of $n$,
the number of zero crossings,
\bea
\log P(p,t)= \sum_{r=1}^{\infty} { (\log p)^r\over {r!}}{\langle
n^r\rangle}_c,
\label{moments}
\eea
where ${\langle n^r\rangle}_c$ are the cumulants of the moments. Using
$p=1-\epsilon$, we express the right hand side as a series in powers of
$\epsilon$. Fortunately the computation of the moments of $n$ is
relatively straightforward, though tedious for higher moments. We have
already mentioned the result of Rice for the first moment. The second
moment $\langle n^2\rangle$ was computed by Bendat\cite{bendat}. We have 
computed the third moment as well\cite{majbray}. For example, for $2$-d
diffusion equation, we get the series,
\bea
\theta(p=1-\epsilon)={1\over {2\pi}}\epsilon +\big ( {1\over
{\pi^2}}-{1\over {4\pi}}\big ){\epsilon}^2 +O(\epsilon^3).
\label{series}
\eea
Keeping terms up to second order and putting $\epsilon=1$ (in the same
spirit as $\epsilon$ expansion in critical phenomena) gives,
$\theta(0)=(\pi+4)/{4\pi^2}=0.180899\ldots$, just $3.5\%$ below the
simulation value, $\theta_{sim}=0.1875\pm 0.001$. This thus gives us a
systematic series expansion approach for calculating the persistence
exponent for any smooth Gaussian process.

Note that both the above approaches (IIA and series expansion) are valid 
only for ``smooth" Gaussian processes ($\alpha=2$) with finite density
$\rho$ of zero crossings. What about the
nonsmooth processes where $0<\alpha <2$, where such approaches fail?
Even the Markov process, for which $f(T)=\exp (-\lambda T)$ is a
non-smooth process with $\alpha=1$. Fortunately however for the Markov
case, one knows that the persistence exponent $\theta=\lambda$ exactly.
One expects therefore that for Gaussian processes which may be nonsmooth
but ``close" to a Markov process, it may be possible to compute $\theta$
by perturbing around the Markov result.

In order to achieve this, we note that the persistence $P_0(T)$ in
stationary time $T$, can be written formally\cite{majsire} as the ratio of
two path
integrals, 
\bea
P_0(T)={{2\int_{\phi>0}{\cal D}\phi(\tau)\exp[-S]}\over { \int {\cal
D}\phi(\tau) \exp [-S]} }= {Z_1\over {Z_0}}
\label{part}
\eea
where $Z_1$ denotes the total weight of all paths which never crossed
zero,
i.e., paths restricted to either positive or negative (which accounts for
the factor $2$) side of $\phi=0$ and $Z_0$ denotes the weight of all paths
completely unrestricted. Here $S={1\over {2}}\int_0^T\int_0^T
\phi(\tau_1)G(\tau_1-\tau_2)\phi(\tau_2)d\tau_1d\tau_2$ is the ``action" with
$G(\tau_1-\tau_2)$ being the inverse matrix of the Gaussian correlator
$f(\tau_1-\tau_2)$. Since, $P_0(T)$ is expected to decay as $\exp (-\theta
T)$ for large $T$, we get,
\bea
\theta= -\lim_{T\to \infty} {1\over {T}}\log P_0(T).
\label{energy}
\eea
If we now interpret the time $T$ as inverse temperature $\beta$, then 
$\theta=E_1-E_0$ where $E_1$ and $E_0$ are respectively the ground states
of two ``quantum" problems, one with a ``hard" wall at the origin and the
other without the wall.

For concreteness, first consider the Markov process, $f(T)=\exp(-\lambda
|T|)$. In this case, it is easy to see that $S$ is the action
of a harmonic oscillator with frequency $\lambda$. The ground
state energy, $E_0=\lambda/2$ for an unrestricted oscillator with
frequency $\lambda$. Whereas, for an oscillator with a ``hard" wall at the
origin, it is well known that $E_1=3\lambda/2$. This then reproduces the
Markovian result, $\theta=E_1-E_0=\lambda$. For processes close to Markov
process, such that $f(T)=\exp (-\lambda T) +\epsilon f_1(T)$, where
$\epsilon$ is small, it is then straightforward to carry out a
perturbation expansion around the harmonic oscillator action in orders of
$\epsilon$\cite{majsire}. The exponent $\theta$, to order $\epsilon$, can
be expressed as,
\bea
\theta= \lambda\big (1-\epsilon {2\lambda\over
{\pi}}\int_0^{\infty}f_1(T)[1-\exp (-2\lambda T)]^{-3/2}dT\big ).
\label{perturb}
\eea

At this point, we go back momentarily to the zero temperature Glauber
dynamics of Ising model. Note that the spin at a site in the Ising model
takes values either $1$ or $-1$ at any given time. Therefore, one really
cannot consider the single site process $s(t)$ as a Gaussian process.
However one can make a useful approximation in order to make contact with
the Gaussian processes discussed so far. This is achieved by the so called
``Gaussian closure" approximation, first used by Mazenko\cite{mazenko} in
the context of phase ordering kinetics. The idea is to write, $s(t)=sgn[
\phi(t)]$ where $\phi(t)$ now is assumed to be Gaussian. 
This is clearly an
approximation. However, for phase ordering kinetics with nonconserved
order parameter, this approximation has been quite accurate\cite{mazenko}.
Note that, within this
approximation, the persistence or no flipping probability of the Ising
spin $s(t)$ is same as the no zero crossing probability of the underlying
Gaussian process $\phi(t)$.
Assuming $\phi(t)$ to be Gaussian process, one can compute its
two-point non-stationary correlator self-consistently. Then, using the
same ``log-time" transformation (with $T=\log(t)$) mentioned earlier,
one can evaluate the corresponding stationary correlator $f(T)$.
We are thus back to the general problem of
zero crossing of a GSP even for the Ising case, though only
approximately. 

In $1$ dimension, the correlator $f(T)$ of the underlying process can be
computed exactly, $f(T)=\sqrt {2/(1+\exp(2|T|)}$\cite{majsire} and in
higher dimensions, it can be obtained numerically as the solution of a
closed differential equation. By expanding around, $T=0$, we find that in
all dimensions, $\alpha=1$ and hence they represent non-smooth processes with
infinite density of zero crossings. Hence we can not use IIA or series
expansion result for $\theta$. Also due to the lack of a small
parameter, we can not think of this process as ``close" to a Markov
process and hence can not use the perturbation result. However, since
$\theta=E_1-E_0$ quite generally and since $\alpha=1$, we can use a
variational approximation to estimate $E_1$ and $E_0$. We use as trial
Hamiltonian that of a harmonic oscillator whose frequency $\lambda$ is our
tunable variational parameter\cite{majsire}. We just mention the results
here, the details can be found in \cite{majsire}. For example, in $d=1$,
we find $\theta\approx 0.35$ as compared to the exact result
$\theta=3/8$. In $d=2$ and $3$, we find $\theta \approx 0.195$ and $0.156$. The
exponent in $2$-d has 
recently been measured experimentally\cite{yurke} in a liquid crystal system which has an
effective Glauber dynamics and is in good
agreement with our variational prediction. 

So far we have been discussing about the persistence of a single spin in
the Ising model. This can be immediately generalized to the persistence of
``global" order parameter in the Ising model\cite{crit}. For example, what
is the probability that the total magnetization (sum of all the spins)
does not change sign upto time $t$ in the Ising model? It turns out that
when quenched to zero temperature,
this probability also decays as a power law $\sim t^{-\theta_g}$ with an
exponent $\theta_g$ that is different from the single spin persistence
exponent $\theta$. For example, in $1$-d, $\theta_g=1/4$
exactly\cite{crit} as opposed to $\theta=3/8$\cite{derrida2}. 
A natural interpolation between the local and global persistence can be
established
via introducing the idea of ``block" persistence\cite{clement}. The
``block" persistence is the probability $p_l(t)$ that a block of
size $l$ does not flip upto time $t$. As $l$ increases from $0$ to
$\infty$, the exponent crosses over from its ``local" value $\theta$ to
its ``global" value $\theta_g$.

When quenched to the critical temperature $T_c$ of the Ising model, the
local persistence decays exponentially with time due to the flips induced
by thermal fluctuations but the ``global" persistence still decays
algebraically, $\sim t^{-\theta_c}$ where the exponent $\theta_c$ is a
new non-equilibrium critical exponent\cite{crit}. It has been computed in
mean field theory, in the $n\to \infty$ limit of the $O(n)$ model, to first
order in $\epsilon=4-d$ expansion\cite{crit}. Recently this epsilon
expansion
has been carried out to order $\epsilon^2$\cite{klaus}.  

Recently the persistence of a single spin has also been generalized to
persistence of ``patterns" in the zero temperature dynamics of $1$-d Ising
or more generally $q$-state Potts model.
For example, the survival probability of a given ``domain"
was found to decay algebraically in time as $\sim
t^{-\theta_d}$\cite{krapiv} where the $q$-dependent exponent
$\theta_d(2)\approx
0.126$\cite{krapiv} for $q=2$ (Ising case), different from $\theta=3/8$
and
$\theta_0=1/4$.
Also the probability that a ``domain" wall has not
encountered any other domain wall upto time $t$ was found to decay
as $\sim t^{-\theta_1}$ with yet another new exponent $\theta_1(q)$
where $\theta_1(2)=1/2$ and $\theta_1(3)\approx 0.72$\cite{majcor}.
Thus it seems that there is a whole hierarchy of nontrivial exponents
associated with the decay of persistence of different patterns in phase
ordering systems.

Another direction of generalization has been to investigate the
``residence time" distribution, whose limiting behaviour determines the
persistence exponent\cite{dornic}. Consider the effective single site
stochastic
process $\phi(t)$ discussed in this paper. Let $r(t)$ denote the
fraction of time the process $\phi(t)$ is positive (or negative)
within time windon $[0,t]$. The distribution $f(r,t)$ of the random
variable $r$ is the residence time distribution. In the limits $r\to
0$ or $r\to 1$, this distribution is proportional to usual
persistence. However the full function $f(r,t)$ obviously gives more
detailed information about the process that its limiting behaviours.
This quantity has been studied extensively for diffusion
equation\cite{dornic,newman}, Ising model\cite{dg}, Le'vy
processes\cite{bald}, interface
models\cite{tnd} and generalized Gaussian Markov processes\cite{abhi}. 

The various persistence probabilities in pure systems have recently been
generalized to systems
with disorder\cite{fisher}. For example, what is the probability
that a random
walker in a random environment (such as in Sinai model) does not cross 
the origin? Analytical predictions for the persistence in disordered
environment have been made recently based on an asymptotically exact
renormalization group approach\cite{fisher}. 

Another important application of some of these persistence ideas,
experimentally somewhat more relevant
perhaps, is in the area of interface fluctuations\cite{krug,kallabis}.
The persistence in Gaussian interfaces such as the Edwards-Wilkinson
model, the problem can again be mapped to a general GSP but with a
non-Markovian correlator\cite{krug}. In this case, several upper and lower
bounds have been obtained analytically\cite{krug}. For nonlinear
interfaces of the KPZ types, one has to mostly resort to numerical
means\cite{kallabis}. The study of history dependence via persitence
has provided some deeper insights in the problems of interface
fluctuations\cite{kallabis,tnd}.

On the experimental side, the persistence exponent has been measured in
systems
with breath figures\cite{marcos}, soap bubbles\cite{tam} and twisted
nematic liquid crystal exhibiting planar
Glauber dynamics\cite{yurke}. It has also been noted recently\cite{ronald}
that persistence exponent for diffusion equation may possibly mbe measured
in dense spin-polarized noble gases (Helium-3 and Xenon-129) using NMR
spectroscopy and imaging\cite{tseng}. In these systems the polarization
acts like a diffusing field. With some modifications these systems may
possibly also be used to measure the persistence of ``patterns" discussed
in this paper. 

In conclusion, persistence is an interesting and challenging
problem with many applications in the area of 
nonequilibrium statistical physics. Some aspects of the problem has been
understood recently as reviewed here. But there still exist many questions
and emerging new directions open to more theoretical and experimental 
efforts.  

I thank my collaborators C. Sire, A.J. Bray, S.J. Cornell, J. Krug, H.
Kallabis, B. Yurke, A. Pargellis and A. Dhar. I especially thank D. Dhar
for many valuable suggestions and discussions. I thank M. Barma, B.
Derrida and C. Godr\`eche for useful discussions. I am grateful to CNRS,
Universite' Paul Sabatier for hospitality where the whole series of work
began.

\end{multicols}
 
\end{document}